\documentclass[10pt, conference, letterpaper]{IEEEtran}
\usepackage{graphicx}
\usepackage{hyperref}
\usepackage{pdfpages}
\usepackage{textcomp}
\usepackage{amsmath}
\usepackage{algorithm}
\usepackage[noend]{algpseudocode}
 \usepackage{setspace}
 \usepackage{bm}
\usepackage{mathtools}
\let\Algorithm\algorithm
\renewcommand\algorithm[1][]{\Algorithm[#1]\setstretch{1}}
\newcommand{\M}[1]{\bm{\mathit{#1}}}
\ifCLASSINFOpdf
\else
\fi
\begin{document}
%
\title{Securing Vehicle to Vehicle Communications using Blockchain through Visible Light and Acoustic Side-Channels}


\author{\IEEEauthorblockN{Sean Rowan, Michael Clear, Mario Gerla\IEEEauthorrefmark{1}, Meriel Huggard, Ciar\'an Mc Goldrick}
\IEEEauthorblockA{School of Computer Science and Statistics, Trinity College Dublin, Ireland\\ Department of Computer Science, UCLA, Los Angeles, USA\IEEEauthorrefmark{1} \\ sprowan@tcd.ie, clearm@tcd.ie, gerla@cs.ucla.edu\IEEEauthorrefmark{1}, Meriel.Huggard@cs.tcd.ie, Ciaran.McGoldrick@scss.tcd.ie }}


%


\maketitle

\begin{abstract}
Autonomous and self-driving vehicles are appearing on the public highways. These vehicles commonly use wireless communication techniques for both vehicle-to-vehicle and vehicle-to-infrastructure communications. Manufacturers, regulators and the public are understandably concerned about large-scale systems failure or malicious attack via these wireless vehicular networks. This paper explores the use of sensing and signalling devices that are commonly integrated into modern vehicles for side-channel communication purposes. Visible light (using a CMOS camera) and acoustic (ultrasonic audio) side-channel encoding techniques are proposed, developed and evaluated in this context. The side-channels are examined both theoretically and experimentally and an upper bound on the line code modulation rate that is achievable with these side channel schemes in the vehicular networking context is established. A novel inter-vehicle session key establishment protocol, leveraging both side-channels and a blockchain public key infrastructure, is then presented. In light of the limited channel capacity and the interoperability/security requirements for vehicular communications, techniques for constraining the throughput requirement, providing device independence and validating the location of the intended recipient vehicle, are presented. These reduce the necessary device handshake throughput to 176 bits for creating symmetric encryption and message authentication keys and in verifying a vehicle's certificate with a recognised certification authority.
\end{abstract}


%
\IEEEpeerreviewmaketitle

\section{Introduction}
Vehicular communication systems are becoming more widely adopted as vehicles are given increased autonomy in the world. The response times of electronic systems are much faster than human-in-the-loop control systems, and this is a motivator behind increasing autonomy in vehicles as they can enhance safety. There is significant, and growing, research activity in vehicular communications. A key interest has been early stage autonomous systems that identify and respond to impending threats (e.g. someone about to run a red light) and having this information exchanged in a wireless channel between the vehicles, with the vehicles issuing warning indicators to the human operators or alternatively proactively taking control of the vehicle and braking\cite{platooning2}. These concepts have developed into systems that can maintain more extensive control over the entire vehicle and, subsequently, groups of vehicles; for example, platooning \cite{platoon123}, where one vehicle leads a group of vehicles and spacing between the vehicles is strictly controlled using inter-vehicle communications.  Amongst other benefits, this can help mitigate shockwaves that occur on highways due to human-in-the-loop reaction times after a vehicle brakes\cite{platooning1}.

A significant concern in intervehicle control scenarios is that of security and availability of data. A malicious attack against a vehicle is of great concern due to the fact that the vehicle can be operated entirely autonomously and, potentially, can be used as a weapon\cite{platooning1}. The wireless communications between vehicles and infrastructure are a particular cause for concern as they can be relatively easily intercepted and swamped by a suitably equipped attacker. For example, wireless transmissions in the 2.4 GHz band are not localised to specific vehicular traffic, making it practicable for a mobile attacker to travel around  and propagate malicious attacks and destructive interference against legitimate vehicular communications.  Natural disasters e.g. earthquake, and other emergency scenarios may result in the failure or unavailability of the wireless infrastructure, whilst simultaneously giving rise to significant exodus of vehicles from the region.  

Previous work has proposed the use of hardware related side-channels \cite{platooning1} as a mitigation to this interference and eavesdropping problem in vehicular networking.  Of particular interest, and one of the initial motivators for our work, was the observation that both ultrasonic systems (parking aid) and visual camera technology (adaptive headlight cornering camera) already exist on the front of many modern vehicles.  More extensive camera capabilities are also available through the impact and accident logging devices that are increasingly incorporated in new vehicles. Ultrasonic audio and CMOS camera visual light devices exhibit high levels of directionality, and favourable signal attenuation properties that make it physically more difficult for an attacker to eavesdrop, intercept, inject or generate interference in these channels. A drawback to the use of these ``commodity" hardware side-channels, however, is that they have a relatively low throughput, thus requiring the development and implementation of protocols that minimise the throughput requirement across the channels.

In this paper we explore the feasibility of establishing the secure infrastructure for low data rate intervehicular communications, using sidechannels, in the absence of continuous RF or wireless infrastructure support. 

At a high level, the key contributions of this work can be summarized as follows:
\begin{itemize}
\item This paper presents the \textbf{first implementation and evaluation of the effective use of ultrasonic audio and CMOS camera visual light side-channels for secure inter-vehicle communications}.
\item The hardware and encoding system is described and its performance is evaluated and characterised.
\item The system and its implementation is demonstrated to have \textbf{significant physical and structural security and integrity} properties.
\item The constraints imposed by these side-channels necessitated the development of a \textbf{new scheme for small throughput, secure, attributable exchange of session key information} between vehicles.
\item The \textbf{identity} of the vehicle being communicated with is \textbf{authenticated}, both through certificate checks and visual identity (image processing) means.
\item Blockchain is used as the inter-vehicle message transport, as it provides a \textbf{secure, verifiable, shared, open and distributed ledger}.  It is also device and platform agnostic.
\item System models, algorithmics and select performance evaluations are provided to demonstrate the performance and potential of the scheme.
\end{itemize}  

\section{Hardware Related side-channels}

\begin{figure}[ht]
\vspace{-0.4cm}
  \centering
    \includegraphics[width=0.5\textwidth]{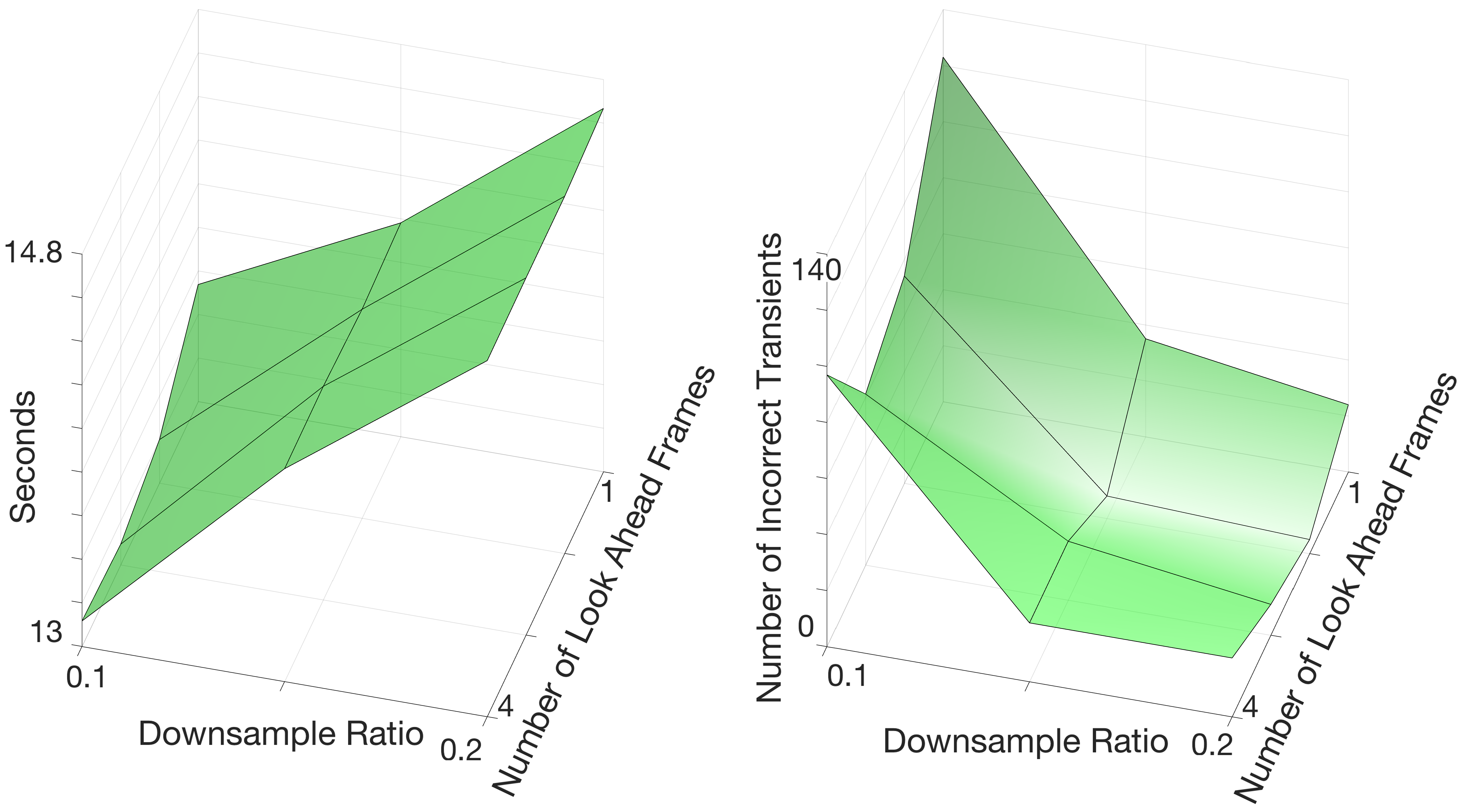}
    \vspace{-0.5cm}
    \caption{This  CMOS camera experiment is a hill-climbing approach to finding optimal software controlled configurations across 8 different exposure times (ranging from 0.1 to $2.2 \times 10^{-11}$ seconds/pixel).  Note that the long processing times are due to file access delays; the subsequent operational characteristics presented will remain valid for real-time, in-memory systems.}
\label{fig2}\end{figure}

Vehicular networking situations present some interesting challenges for effective data throughput using (commodity camera) visual side-channels. In many of the evaluated scenarios, the reliable throughput of the visual side-channels can be as low as 15 bits/s for a 30 frame/s CMOS camera using line-coding techniques. This paper focuses on the use of standard CMOS cameras, of the type currently employed in modern vehicles.  Two specific problems arise in the use of these cameras - those due to transient frames and exposure control.  A transient frame is a captured frame where the transmitted data is not fully captured by the receiver because the modulation time of LED's is much faster than a CMOS camera's exposure time leading to a physical upper bound on the maximum signaling rate.  A transient frame can not be decoded into either a 1 or a 0 if it is taken independently. For robustness, transient frames require a safety margin beyond the Nyquist sampling rate which limits the maximum signalling rate.  A hill-climbing approach to finding optimal software controlled configurations is shown in Fig \ref{fig2}. This involved detecting modulation using the approach described in Fig. \ref{dbd} for each of 8 different exposure times, ranging from 0.1 to $2.2 \times 10^{-11}$ seconds/pixel, with a goal of minimising detection errors and processing time. The value colour channel is used as it is hue independent and therefore can be used for detecting modulation from head or tail lights on a car. The number of transition mistakes metric shows that the best configuration is a high downsample ratio and 2 look ahead frames. Evaluating adjacent frames in a sequence tends to fail due to the transient frame effect. A lookahead value greater than 2 leads to an increase in error as frames become less temporally correlated. Higher frame rates and increases in the captured modulation rate (as described by Danakis in \cite{danakis}) were evaluated but offered little gain due to the outdoor, noisy, occluded and motion-involved scenarios that prevail in our target environment.  Emerging camera technologies offer significant promise for higher throughput and data rates in vehicular scenarios; in particular advances in Visible Light Communication optimised cameras e.g. \cite{goodcmos}, describe experimental CMOS cameras which contains several special pixel areas that react very quickly to rapid changes in incident light in unconstrained environments. These classes of device will benefit future vehicular networking scenarios (in particular side-channel systems) as they offer impressive 15-Mbps/pixel data rates.  Algorithm 1 describes the flow and operation of one technique for line-code signaling detection using a CMOS camera.

\begin{figure}[ht]
  \centering
    \includegraphics[width=0.45\textwidth]{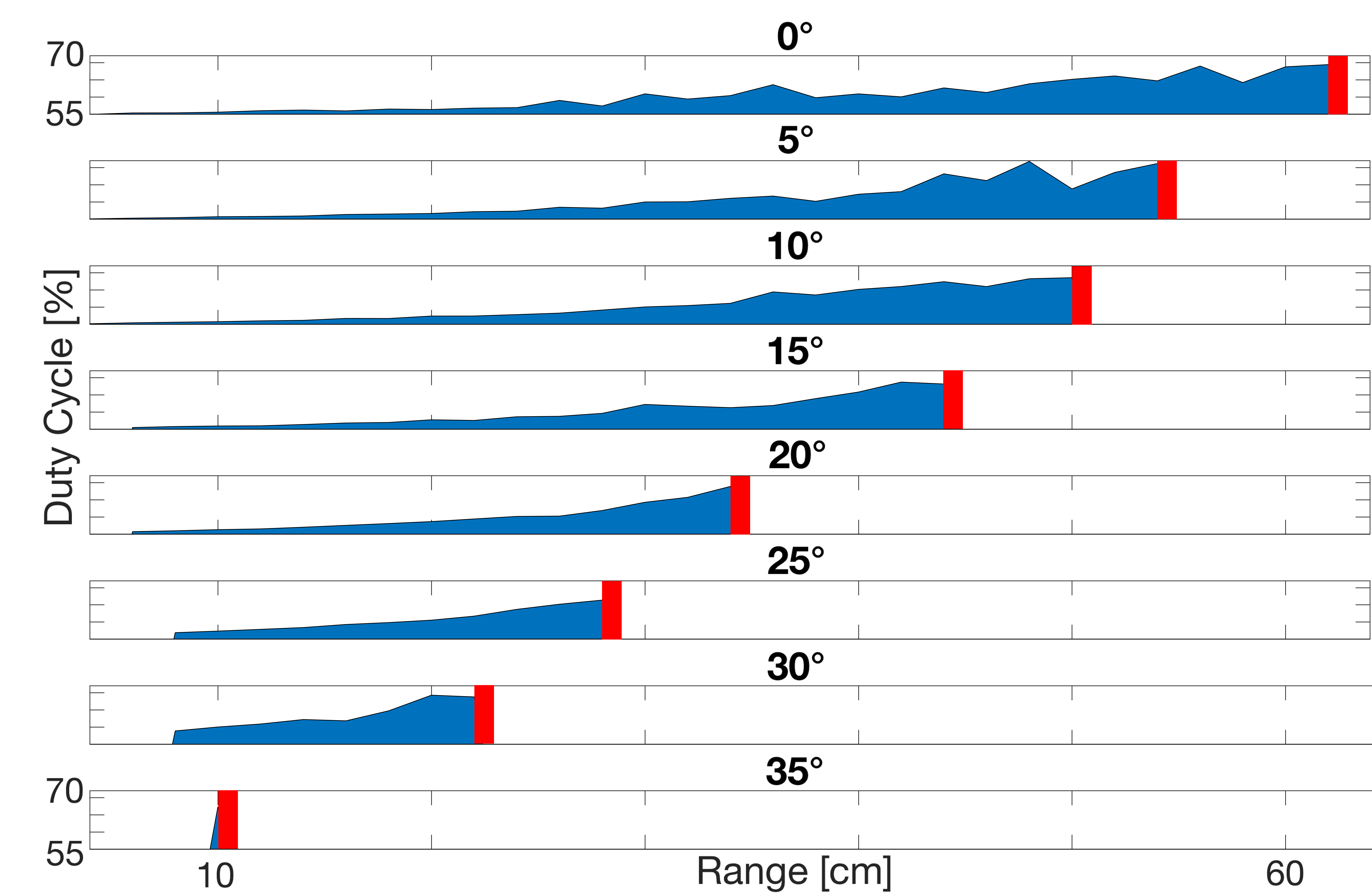}
    \caption{Evaluation of the ultrasonic transducers shows that an increase in duty cycle of the received wave form is correlated with both an increase in angle off center and distance away from the transmitter. The system gain was reduced for ease of testing, with validation performed at common vehicular sensing ranges.}
    \label{fig3}
\end{figure}

\begin{figure}[!ht]
  \centering
    \includegraphics[width=0.4\textwidth]{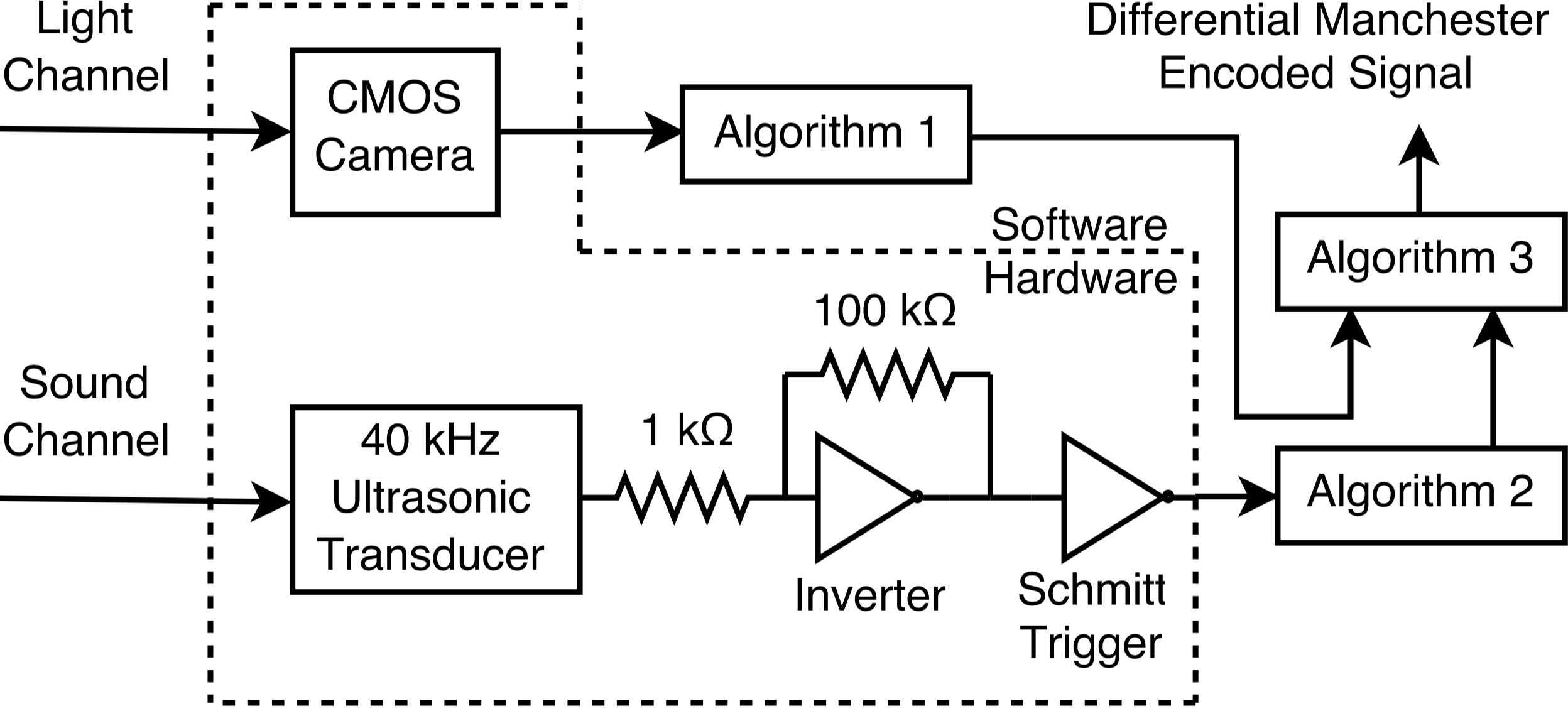}
    \caption{Novel decoding procedure that enables effective reuse of the exterior sensing systems as side channels in a modern vehicle.}
    \label{dbd}
\end{figure}

In commodity form the ultrasonic channel is constrained by the necessity to coexist with the parking aid functionality.  Again, using line coding techniques\cite{manchester}, bidirectional communication was achieved across this side channel at rates of 2Kbps using a 40kHz ultrasonic transceiver.  The usable duty cycle of the ultrasonic channel is, as expected, constrained by both the incident angle of the signal and the transmitter proximity (see Fig. \ref{fig3}.  The period of the received waveform is not affected by these changes. This finding influenced our design decision see Fig. \ref{dbd}) to detect modulation through the measurement of rising edges in the transduced sound wave. Algorithm 2 describes one technique for detecting and decoding an acoustically modulated signal transition.

For both side channels the noisy and occluded environment, caused by two moving vehicles communicating with each other in an ad-hoc bidirectional manner, proved to be a significant limiting factor in establishing the signalling and communication constraints and upper bounds for reliable throughput. Algorithms 3 and 4 describe the binary classifier and Overlap Processing approach used in this work.

The limited usable data rates and constrained channel capacity could be construed as a major drawback to the present-day use of these side-channels for intervehicular communication.  However we recognise that technological and device advances will see rapid and significant increases in the available data rates and channel capacities across both of these side-channels.  Moreover the presence of both ultrasonic and visual camera systems in present-day vehicles argues to both established platform metaphors and also to low cost integration and deployability opportunities for such side channel approaches. Finally the robust and directional nature of both sidechannels affords physical security to the data exchanges, in addition to any signal encoding employed.

\begin{algorithm}
\caption{Video Preprocessing}\label{alg:alg1}
\begin{algorithmic}[1]
\Procedure{VideoPreprocess}{$\M{F}, r, l, a$}\Comment{$\M{F}$ is a 4-dimensional matrix of $n$ RGB image frames gathered from a CMOS camera, $r$ is the frame rate, $l$ is the number of quantisation levels and $a$ is the number of lookahead frames for frame comparison. From experiments, values of $l=256$ and $a=2$ are chosen.}
\State $\M{\Phi} \gets (\M{X},\M{Y},\M{S},\M{\Theta},\M{Z}) \gets \text{LPR}\{\M{F}\}$ \Comment{Location $(\M{X},\M{Y})$, scale $\M{S}$, rotation $\M{\Theta}$ and skew $\M{Z}$ of a detected license plate in the frame.}
\State $\M{F} \gets \text{Quantise}\{\text{Max}\{\M{F}.r,\M{F}.g,\M{F}.b\},l\}\}$ \Comment Convert to the value colour channel and quantise the resulting image.
\State $\M{F} \gets \text{Segment}\{\M{F},\text{T}\{\M{\Phi} \}\}$ \Comment{Segment an elliptical region around the tail/head lights of the vehicle defined by a geometric transformation of $\M{\Phi}$ (see figure \ref{phi}).}
\State $\M{F} \gets \sum_{j=1}^{m}\M{F}[i, j]\big( \frac{1}{\sigma\sqrt{2\pi}}e^{-\frac{1}{2}(\frac{j-\mu}{\sigma})^2}\big)\big/(l*m), \ \ \forall \ i$\Comment{Apply a Gaussian kernel to the elliptical region and normalise.}
\State $\M{F} \gets 1 - \frac{\big(\M{F}[i,:] - \bar{\M{F}}[i,:]\big)\big(\M{F}[i+a,:] - \bar{\M{F}}[i+a,:]\big)^\top}{\sqrt{\big(\M{F}[i,:] - \bar{\M{F}}[i,:]\big)^\top}\sqrt{\big(\M{F}[i+a,:] - \bar{\M{F}}[i+a,:]\big)^\top}}, \ \ \forall \ i < (i -a)$\Comment{Compare frames $a$ spaces apart using the distance correlation metric.}
\State $\M{F} \gets \text{Distance}\{\text{Rising}\{\text{Otsu}\{\M{F}\}\}\}/r$ \Comment{Classify the correlations into two classes (see Otsu\cite{otsu}) and set the matrix equal to the distance between rising edges in seconds.}
\State \textbf{return} $\M{F}$
\EndProcedure
\end{algorithmic}
\end{algorithm}

\begin{figure}[!ht]
  \centering
    \includegraphics[width=0.31\textwidth]{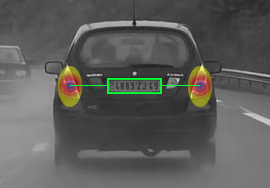}
    \caption{Mapping of $\phi$ (license plate location) to $T\{\phi\}$ (light modulation source) for a car. A gaussian kernel is used at the location of $T\{\phi\}$ to weight the pixels close to the light modulation source, which reduces errors due to relative movements of the vehicles and distortions on the mapping of $T\{\phi\}$ and normalisation is applied to allow the relaxation of dependence of the distance between vehicles.}
    \label{phi}
\end{figure}

\begin{algorithm}
\caption{Sound Preprocessing}\label{alg:alg2}
\begin{algorithmic}[1]
\Procedure{SoundPreprocess}{$v,m$}\Comment{$v$ is a voltage measurement of the ultrasonic audio transducer and $m$ is the modulation period. From experiments, a value of $m$ = 20 ms is chosen based on the processing speed of the embedded device. $d$ is buffered before entering Alg. 3.}
\State \textbf{normal routine:}
\State $v \gets \text{Voltage Measure}\{\}$
\State \textbf{interrupt routine:}
\If{$v \equiv $ \text{rising edge}}
\State $d \gets \text{RisingEdgeDist}\{\}$ \Comment{Distance in seconds between this rising edge and the previous rising edge.}
\If{$d > m/2$}
\State \textbf{return} $d$
\EndIf
\EndIf
\EndProcedure
\end{algorithmic}
\end{algorithm}

\begin{algorithm}
\caption{Overlap Processing}\label{alg:alg3}
\begin{algorithmic}[1]
\Procedure{OverlapProcess}{$\M{F}$}\Comment{Obtain the differential manchester encoded signal.}
\State $\mu \gets j \gets 1$
\State $\M{F} \gets \text{Otsu}\{{\M{F}}\}$
\For{$1 \leq i \leq \text{Size}\{\M{F}\}$}
\If{$\M{F}[i] = 1$}
\State $\M{G}[j] \gets \M{G}[j+1] \gets \mu$
\State $j \gets j + 2$
\Else
\State $\M{G}[j] \gets \mu$
\State $j \gets j + 1$
\EndIf
\State $\mu \gets \mu \oplus 1$
\EndFor
\State \textbf{return} $\M{G}$
\EndProcedure
\end{algorithmic}
\end{algorithm}


\subsection{Physical security}
The side channel approach accrues both physical and structural security properties.  The ultrasonic channel has directional propagation characteristic, in consort with the close range of the platooning vehicles (typically no more than a few metres) \cite{platooning2}.  These sensors are bumper mounted and positioned to detect obstacles directly aligned with, or immediately to the side of, the vehicle bumper.  Their ranging properties are typically tuned to a {\raise.17ex\hbox{$\scriptstyle\sim$}}10ft range and a $\pm$45\textdegree \ detection angle.  Intercepting or injecting data into the bidirectional data exchange necessitates positioning a transceiver in sight of the ultrasonic transponders on both vehicles, in circumstances where the platooning vehicles may be traveling at more than 60mph.
On first consideration the visual channel may appear to be more vulnerable to malicious external signalling.  Vehicular cameras often have lens with a wide field of view, and headlight cameras are positioned towards the extremities of the vehicle where they may be more susceptible to unwanted targeting.  Lasers and directive light sources can also be used to attempt to ``blind'' a device by overloading its sensor.  In experiments these problems have been successfully surmounted using structural security properties.  By this it is meant that camera processing is only targeted at those areas where it is known that data can be expected.  Specifically, the camera sidechannel logic identifies brake lights on the forward vehicle (1, 2, or 3) at first contact, and validates their presence and functionality.  The dual camera imagery is processed upon receipt in order to identify the visual sidechannel areas i.e. the brake light signals, and all other camera imagery data is disregarded.  In general, the structural properties of the sidechannel signalling areas remain unchanged; even through cornering and acceleration/deceleration behaviours.  A change in the visible forward vehicle e.g. because of a lane change, necessitates a reidentification process.  It is important to note that the visual channel can also be used to extract vehicle specific identification information e.g. license plate data and other identifying marks and shapes.  Cumulatively the camera sidechannel(s) can provide considerable physical and structural assurance as to the source of the visual imaging data. 

The preceding sections describe the creation and validation of a secure intervehicle communication carrier that exploits the relatively high bit rate and the observable source angle properties of the sidechannels.  The ultrasonic signal can arrive from almost any angle, subject to the physical positioning of the vehicles, and has a `relatively' high bit rate; the visual light signal arrives from an observable angle and has a relatively low bit rate. These properties are exploited to enhance throughput and the overall security of the communications.  Whilst the sidechannel data rates may appear low, particularly when compared with DSRC and wireless approaches in the literature\cite{dsrc}, this scheme can function entirely independently and autonomously from any highly available infrastructural dependence e.g. WiFi, DSRC, or LTE.  During wireless network outages, in the presence of attack or compromise, or for validation of data received through other channels, the sidechannel approach can be successfully employed to keep vehicles and platoons safely controlled and operational.

The following sections set out how the sidechannel system is used to securely establish and disseminate PKI key exchanges using a distributed blockchain implementation between moving (platooning) vehicles.   

\section{Minimising Throughput Requirements In Vehicular Communications Using side-channels}
Communication systems seek to maximize the amount of data encoded and successfully conveyed per transmission interval.  When utilising the CMOS visual light and ultrasonic side-channels, we seek to minimise the required information content, or payload, of the message due to the low throughput capability of the channels.

Many application layer protocols  have been proposed and implemented for inter vehicle communications\cite{vproto}. For providing security and authentication capabilities between nodes\footnote{
A key exchange algorithm usually requires the establishment of a pre-master secret in order to securely construct a master secret. Typically \cite{cryptography}, this is performed by either exchanging fixed RSA public keys and encrypting/authenticating/sending the pre-master secret, exchanging fixed Diffie-Hellman public keys and computing the shared secret directly or exchanging both fixed RSA public keys and authenticated one-time-use Diffie-Hellman public keys to compute a shared secret that is perfectly forward secure \cite{pfs} (which means that an attacker can not decrypt historic messages between machines if the fixed RSA private key is compromised). All three of these exchanges require at least one public key value to be exchanged, which is typically at or above 32 bytes. For authentication by some central authority, a certificate (e.g. X.509 certificate \cite{x509}) might also be exchanged that authenticates the sender's provided public key value, and this certificate is typically 1 kB in length. Key establishment protocols also include random values that are exchanged by both parties in order to introduce randomness into the master secret and protect against duplicated master secrets; these values are typically at least 16 bytes in length.}, these schemes commonly establish symmetric encryption keys, message authentication keys and also perform certificate checks with central certification authorities.
 
In experiments, transmitting worst case key and check data combinations through the side-channels has taken some seconds.  This magnitude of delay is not tractable in a vehicular networking scenario. 
A number of approaches were evaluated in order to reduce the size and volume of necessary exchanges for enabling secure communication key exchanges between the vehicles.  The initial premise was to encapsulate both the public key value and the certificate for every vehicle into a database, such that only a small primary key value was required to map to the required data. This could be naively done online by contacting the ``cloud" and requesting the credentials mid-handshake. However, this approach requires internet connectivity and also creates a broad new attack surface that is easily targeted. 
A superior approach would be to have the entire database offline and travel with the vehicles themselves. One mechanism for achieving this would be to have a centralised database that is secure and widely trusted, and that can be synced to/by the vehicles periodically. It is probable that such a centralised system would not find favour due to the multitude of manufacturers, political regions and vehicle types that may exist in such a downloadable database. 
Blockchain based domain name systems/public key infrastructures \cite{blockstack} were identified as conceptually suitable for our work as they do not require centralisation in order to push data into the distributed database - the integrity of the data is assured by an underlying blockchain.  The desirable principles and features of these technologies for our side-channel system are now articulated.

\subsection{Blockchain Public Key Infrastructure}
\label{economic}
\subsubsection{Blockchain}

A blockchain \cite{bitcoin} is a distributed public ledger that allows a contributor machine to maintain a verifiable record of transactions which are recorded in the form of a series of interconnected blocks of data in the ledger, and are considered to be secure from tampering and revision.  Each node can create and verify blocks and send new transactions to the blockchain.  Blocks hold timestamped batches of transactions or, in some implementations, programs or other data.  Each block includes the hash of the prior block in the blockchain, thereby linking the two blocks.
Mining is the term used to refer to the distributed computational review process performed on each block of data.  The creation of a new block is both computationally costly and time consuming (see Fig. \ref{network_diagram}). When the blockchain is robust, nodes on the network will independently verify successful hashes on the blockchain network and achieve consensus on creation of the next block (even where neither party knows or trusts each other).  For instance, in cryptocurrencies\cite{bitcoin} mining of a block is subject to a difficulty metric and a reward mechanism. The difficulty metric increases as the rate of block generation increases, thereby slowing the block generation rate and mitigating against malicious mining.  When a block is discovered, the discoverer may award themselves a reward.  In cryptocurrencies, the reward is usually  financial, e.g. a certain number of BitCoins\footnote{Bitcoin was chosen as the underlying blockchain by the Blockstack developers because it has the lowest probability of the 50\% hashing power attack\cite{bitcoin} succeeding (due to having the largest number of nodes participating in the network). The developers provide a mechanism to transition Blockstack over to a different blockchain in the event that one becomes more robust than Bitcoin. Alternatives to Blockstack include NameCoin \cite{namecoin}, which had a dedicated blockchain. However the system did not have sufficient network scale which resulted in a user achieving more than 50\% of the hashing power on the network and demonstrating power attack capacity. Emercoin \cite{emercoin}, is explicitly set up as a blockchain public key infrastructure, where the blockchain is exclusively used for Emercoin. The likely susceptibility of Emercoin to a similar power attack to that of NameCoin motivates our use of the more widely used, larger user base Blockstack system.}, and agreed by everyone on the network.  This is an important facet of the underpinning design of blockchain networks - this reward system makes it economically futile for an attacker to attack the blockchain network in order to gain benefit (i.e. rewrite previous blocks) because the cost of successfully doing so is beyond the available reward. It is therefore more profitable for nodes to concentrate their computational resources on legitimate mining of hashes in the blockchain because they will increase trust in the system and accrue a higher economic reward in doing so. 
It is of note that nodes cannot collude with each other in the generation of hashes, as the proof that a node correctly made a hash is unique to that node.  This property adds security to the blockchain network by thwarting large scale malicious attacks from compromising the blockchain network. As we propose to use blockchain in widely distributed, open, intervehicular communication scenarios, it is critical that the primary key access encoding infrastructure is demonstrably robust and secure to large scale attack surfaces.  The blockchain infrastructure provides these capabilities.

\subsubsection{Blockstack Decentralised Domain Name Service}
Blockstack is an open source, peer reviewed application stack implementation that provides  identity, naming, storage, and authentication services \cite{blockstack}.  Blockstack allows relatively large amounts of data (max 8kB per entry) to be securely stored in a distributed hash table by storing hash values of the data in robust blockchain blocks (see Figure 4 in \cite{blockstack} for illustration). It can be viewed as a distributed database that does not require central management; attackers cannot amend entries in the distributed hash table without being detected through the checking of the underlying blockchain. A Blockstack server, maintaining its own copy of the distributed hash table, was implemented on a Cloud virtual machine and extensively evaluated.  
Overall, the Blockstack system was found to be very reliable. 

A key feature of the Blockstack system, when used in our vehicular networking scenarios, is that machines can directly download a copy of the distributed hash table from an untrusted source and independently verify the contents using the blockchain with minimal computational requirement \cite{blockstack-bootstrap}. Thus each vehicle in a platoon can independently and individually acquire the required data structure.   There remains scope for further optimisation and evaluation of the timing overhead for hash table lookups on various embedded devices which contain a full copy of the distributed hash table. 


\subsubsection{Economic Benefit of an Attack on a Blockchain When It Contains Extraneous Information}

In section \ref{economic} we note that there can be economic 
It is reasonable to surmise that attacks on blockchain structures may be attempted on vehicular network data or vehicular platoons e.g. by well resourced malicious groupings or foreign powers.
Use of blockchain for intervehicle data interchange is most likely to be, and remain, secure when the userbase is large, thereby mitigating against the 50\% hashing power attack compromise \cite{bitcoin}. 
\footnote{When new information is inserted into the Blockstack distributed hash table, a clever mechanism is used to prevent revision of data as it is being propagated throughout the distributed network of Blockstack servers: a hash of the data about to be inserted to the distributed hash table is first released to the network. After a period of time elapses (currently about 12 hours) the actual data is released to the distributed hash table and its historic hash (existing at a lower block height) can be checked independently. As time progresses and the block height in the blockchain increases, it becomes harder for an attacker to overturn previous block values until  it becomes practically infeasible. In the vehicular networking case, where human life may be threatened, an  elapsed duration to release should likely be of the order of weeks or months.  It remains an interesting, and rather fraught, research question.}  

In the system described herein, the data inserted into the blockchain relates to the PKI key data used to secure intervehicular communications.  As this data can, and will, expire in the highly mobile vehicular environment, and as  the use of physical side-channels for transmission of these blockchain entries is targeted, we express considerable confidence that the system as described is secure under normal usage scenarios.


\subsection{Index Key Size}
\begin{figure}[!ht]
  \centering
    \includegraphics[width=0.4\textwidth]{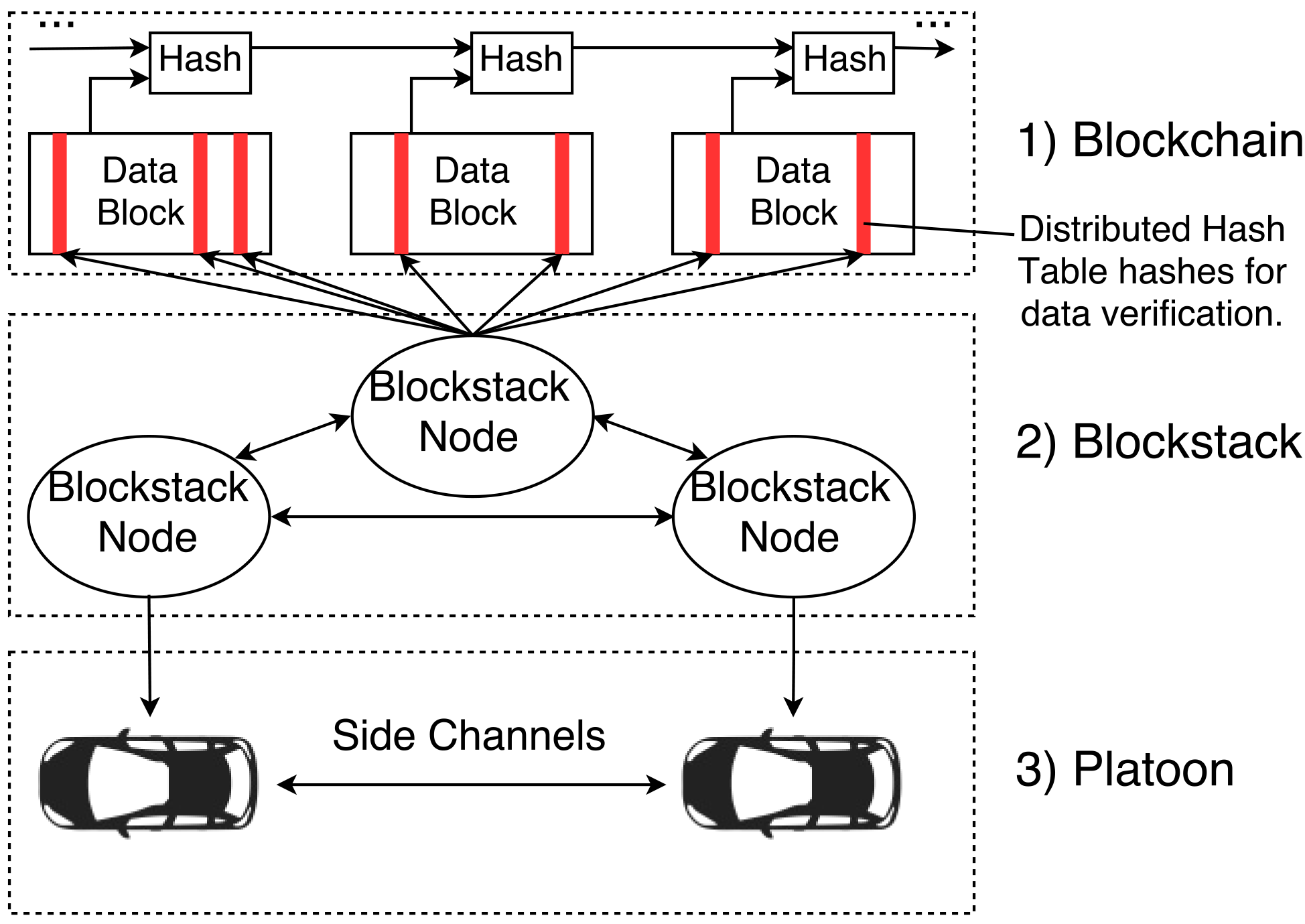}
    \caption{Network diagram of a blockchain-based public key infrastructure for a vehicular platoon.   1) Blockchain (ledger): Provides underlying security to information contained in Blockstack nodes. 2) Blockstack: Each node maintains its own copy of a distributed hash table containing a public key infrastructure used to authenticate communications between vehicles in a platoon. The data stored is a public key, a Hash\{identity salt $||$ license plate\} and a certificate binding these two values for every vehicle.  3) Platoon: Vehicles in the platoon hold their own copy of the public key infrastructure contained in the Blockstack and securely communicate with other vehicles in the platoon using the physical side channels.  The system works with and without live online connectivity to the blockstack.  Intervehicular symmetric session key generation is described in Figure \ref{sessionestab}. }
    \label{network_diagram}
\end{figure}

A primary key that maps into a database must be unique for every entity that stores information in that database. For vehicles, a useful 'unique' value that is already assigned is the license plate number. It is known to the registered vehicle and can be read, e.g. using OCR, by camera from a following vehicle in a platoon.  As previously noted any data can be inserted into the database (blockchain) without the need for a trusted certificate authority.  As license plate numbers are deterministic, an attacker could ``brick" (i.e. make unusable) license plate numbers in the database by registering plate number ranges before the rightful owner does. Combining the license plate value with a random value ``salt", known only to the generating vehicle, immediately prior to the generation of the public key information for the exchange means that an attacker must seek to register all license plate + salt combinations in order to succeed with a ``brick" attack.  Ideally the random salt size would be as small as possible as this data, of a fixed length (see Fig. \ref{network_diagram}), will be transmitted through the vehicular side-channels when creating the database primary key.  
The primary key value can be of a fixed size by calculating it using a hash function as follows:\\

primary key $\gets$ Hash\big\{license plate value $||$ identity salt\big\}\\

A reasonable key dimension, combined with a ``cost" to insert data into the block chain, provide an effective deterrent to this attack approach. 

Inserting data into a blockchain, and therefore a blockchain distributed hash table (such as Blockstack) is not free. As the scheme presented herein exploits the scale and number of users on BitCoin to provide security and robustness against the 50\% hashing power attack surface, transactions on the blockchain network requires the payment of small miner's fees as a reward mechanism for nodes to validate new blocks in the network. Currently, this cost is about 0.05 USD. The Blockstack mechanism also has its own additional fees built in that brings the entire cost of a single low volume insertion to 0.09 USD. Thus a brick attack against a license plate with a salt will incur a significant financial cost.  For a 32 bit salt length, the legitimate license plate owner will pay 0.09USD, whereas the cost of bricking the license plate is $0.09 \times 2^{32}$ or about 400MUSD. This figure varies depending on the value of Bitcoin. We believe that this is a sufficient disincentive to license plate attacks for all but state security interests.  

In any event further efficiency, robustness and reduction in storage and data transmission requirements are gained from the observation that the certificate authority is required to construct the trusted binding between a primary key value and a cryptographic public key value.  When downloading information from the distributed hash table, data can be selected to be downloaded only where entries are associated with certificates from widely known certificate authorities - or the subset of certification authority(ies) that we choose to accept for our vehicular networking exchanges.  Erroneous or malicious insertions can be checked for just before registering with a certificate authority to avoid collisions in the distributed hash table.


As only a single 176 bit data exchange is needed,  the side channel system set up can be achieved quickly and efficiently. The short message size enables rapid symmetric key information exchange. This is essential due to the proximity requirement for the side channel system to function.  In principle,  symmetric key establishment for a platoon has no defined upper time bound; however, in the side channel situation there are both physical and safety constraints on the upper bound.  In practice, the normal sidechannel operational range bounds can be relaxed as the camera system can visually detect vehicles at extended distances and the ultrasonic range can be increased through power adjustment.  These extended ranges are necessary during initial platoon formation when using side channels, during slow start up and in enabling side channel only platooning in speed banded lanes, for example car pool lanes that have a minimum speed.  Clearly the extended range  provision is necessary to maximise intervehicular distance for safe operation  during the exchange of the key data (which leads to synchronised control).  The effective intervehicular distance that, for safety reasons, must be maintained during the initial connection setup will be bounded by the responsivity of the side channel system (e.g. camera, ultrasonics, system controller),  the control and actuation characteristics of both vehicles (e.g. braking, acceleration, steering)  and  regulatory constraints (e.g.  minimum required speed in carpool lanes).


\section{Session Key Establishment Protocol Using side-channels and a Blockchain Public Key Infrastructure}

We now present a novel session key establishment approach designed to integrate the security and communications requirements identified in the previous sections. This protocol adopts several concepts from the TLS 1.2 handshake methodology\cite{tls}. TLS was chosen as a good model to follow because it is the most widely used key exchange protocol on the internet and has withstood attacks for 20 years through various versions, meaning that it is very robust. The main differences between this protocol and the TLS 1.2 handshake are the use of the blockchain distributed hash table to minimise the throughput requirement and the use of the visual light channel incorporating with the visual identifier to strongly verify the location of the vehicle being communicated to. 

The main communications/security features of this protocol are:

\begin{itemize}
\item \textbf{Very small throughput requirement.} The ultrasonic channel requires only 160 bits and the visual light channel only requires 16 bits of data to be exchanged in the handshake.
\item \textbf{Location of the vehicle being communicated to is strongly verified.} The mapping from $\phi$ to $T\{\phi\}$, as shown in figure \ref{phi}, links the vehicle's visual identifier, e.g. license plate, with the information that is contained in the visual light channel. Both pieces of information are required for a successful handshake. The information contained in the visual light channel is elaborated upon below. License plate recognition is a widely researched topic and an advanced license plate recognition system such as OpenALPR \cite{openalpr} may be used to establish the license plate data.
\item \textbf{Vehicle authentication.} A certificate is checked as issued by a widely known/accepted certificate authority to authenticate the vehicle being communicated to.
\item \textbf{Practical forward secrecy.} This algorithm is not perfectly forward secret because it does not feature one-time use Diffie-Hellman cryptographic parameters. However, the 256 bits of random values are exchanged by physical sidechannel causing them to be unavailable to most attack methodologies.  The necessity for perfect forward secrecy depends somewhat on the application in question; control commands between vehicles are not typically considered to be sensitive after the control has been safely executed by the vehicle, as an attacker may also observe the physical response of the vehicles e.g. turning. 
\item \textbf{Secure symmetric encryption and message authentication.}
The symmetric key exchange and set up is achieved very rapidly using only 176 bits. 
\item \textbf{Verification of correct master secret generation.} This information is conveyed in the visual light channel as shown below.
\item \textbf{Type of vehicle being communicated to is (weakly) verified.} Using just the mapping from $\phi$ to $T\{\phi\}$, it becomes much harder for smaller vehicles to impersonate larger vehicles. 
Incorporating computer vision recognition algorithms, e.g. the SIFT algorithm \cite{sift}, to detect scale invariant features typical of certain types of vehicles provides significant additional visual validation capabilities e.g. mitigation of vehicle impersonation attacks. It does, however, add some computational complexity and broaden an existing attack surface so its use is bounded.

\begin{table*}[htb]
  \centering
\footnotesize
  \begin{tabular}{llll}
  \textbf{Variable} & \textbf{Description} & \textbf{Variable} & \textbf{Description} \\ \hline
$I$ & Identity salt, 32 bits (fixed) & $L$ & Visual identifier, (fixed) \\
$\phi \gets (x,y,s,\theta,\zeta)$ & Location $(x,y)$, scale $(s)$, rotation $(\theta)$ and
& $T\{\phi\}$ & Geometric transformation of $\phi$ that points to\\
& skew $(\zeta)$ of origin of $L$ &  & light source origin\\
$Prim$ & Primary key that maps into blockchain DHT &  $S$ & Elliptic curve Diffie-Hellman shared secret\\
& (fixed) & & (different for every vehicle combination)\\
$Pub$ & Curve25519 ECDH public key (fixed) & $Priv$ & Curve25519 ECDH private key (fixed)\\
$R$ & Session salt, 128 bits (changes every session) & 
$C$ & Certificate for $L$, $I$ and $Pub$ (fixed) \\
$K$ & Symmetric encryption key (changes every & $M$ & Message authentication key (changes every\\
& session & & session)\\ 
\textbf{Function}  & & \textbf{Function}  & \\ \hline
SHA256 & Hash function [5] &Curve25519 & Shared secret calculation [2]\\
Blockchain & Distributed hash table lookup [4][3] & 
Scrypt & Password based key derivation function [1]\\
\end{tabular}
\end{table*}
\normalsize
\end{itemize}
In broad summary, this paper establishes a means by which secure communication exchanges between vehicles can be achieved across visual and auditory sidechannels.  The scheme securely establishes the symmetric key for intervehicular communication in the absence of any wireless or other centralised infrastructure.


\begin{figure*}[!ht]
 \vspace{-0.5cm} 
 \centering
    \includegraphics[width=0.82\textwidth]{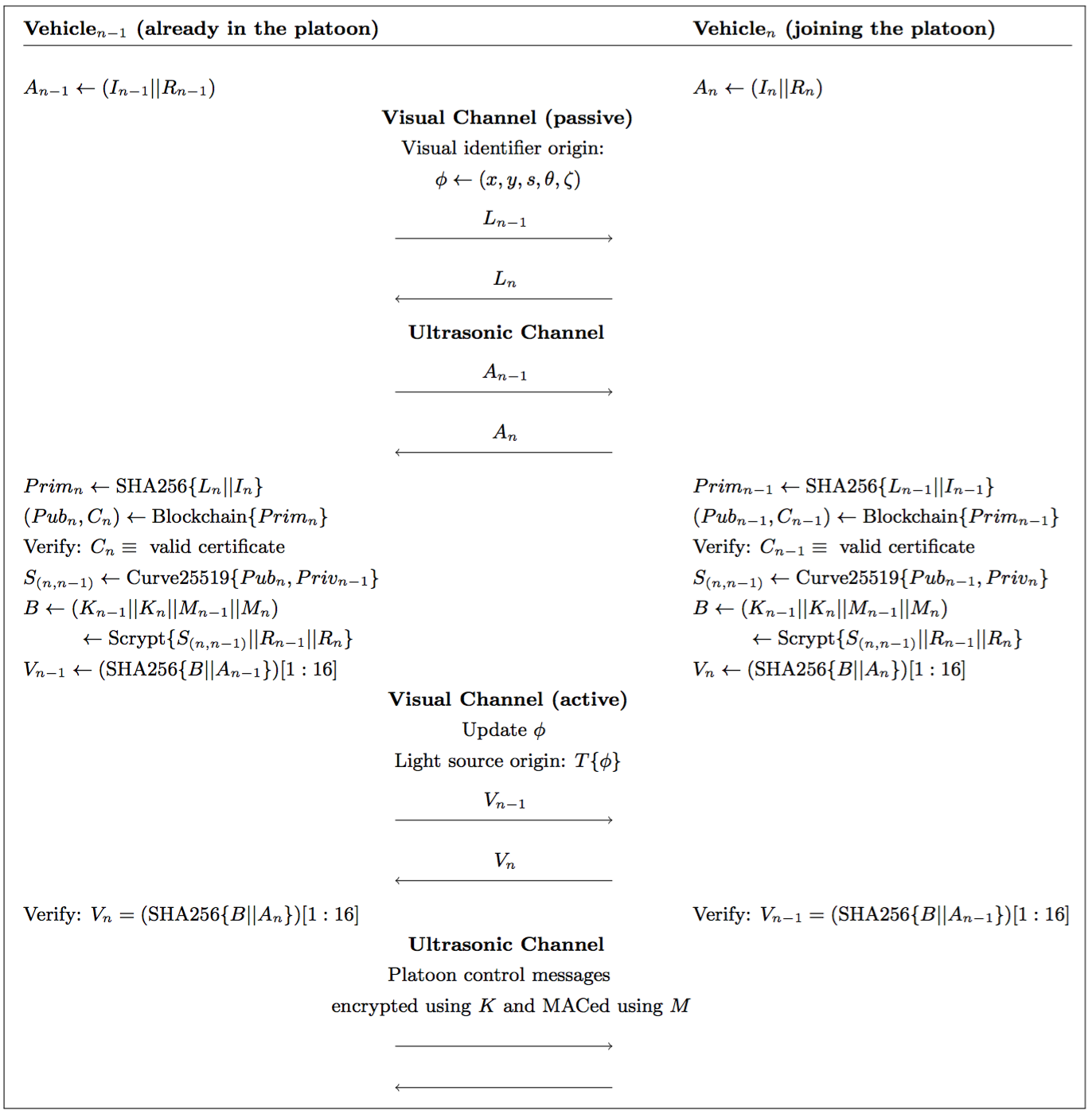}
 \vspace{-0.2cm}  
   \caption{Session establishment protocol using side channels}
    \vspace{-0.5cm}
    \label{sessionestab}
\end{figure*}

\section{Inherent security features of the system}
Beyond the previously described security and robustness features of the scheme and, in particular, the benefits accruing through use of a blockchain based approach, the system will successfully protect against, and fold, a variety of physical and signalling attacks.  For instance:
\begin{itemize}
\item \textbf{Physical attacks to the visual-acoustic channel} e.g. using drones or lasers. Such attacks are folded through exploiting the duality and directionality of the visual/acoustic signals.
\item \textbf{RF jamming (physical) attack, where the symmetric key has survived}.  The visual-acoustic communication side channels remain both intact and fully functional so the vehicular activity, e.g. platooning, can continue.
\item \textbf{RF channel jammed and key compromised.}  Detection of, and protection against, this attack is offered by the integration of blockchain with the sidechannels. 
\end{itemize}

\section{Challenges}
The loss of use of one or more of the side-channels is unavoidable from time to time.  For instance, a vehicle may be involved in a minor accident that breaks a headlight, i.e. a visual compromise, or deforms a bumper, i.e. an ultrasonic compromise. Attackers may also disable channels on purpose;  for example, by breaking a headlight or covering a transducer.  Noting that this work focuses on the use of visual and acoustic carriers as sidechannels, it is anticipated that primary exchanges can, and will, be effected by standard radio frequency transmissions under normal circumstances e.g. DSRC.  Use of the sidechannels as primary key exchange paths would be expected when RF comms are unavailable, when compromise is possible or suspected, or when additional security is required for key exchanges.  For instance, one could envisage high-value convoys primarily employing the sidechannel approach for securing intervehicular exchanges.  In practice, care should be taken to balance usability and security in future implementations of this scheme; for high security settings our recommendation is that key communications take place primarily in the side-channels because attacks on the side-channels are much harder to execute, and more readily detected, than attacks on the standard radio frequencies used in vehicular communications.

\section{Future Work}
\begin{itemize}
\item \textbf{Implementing the side-channels on dedicated embedded devices and assessing the performance improvement in real vehicular communications scenarios}. FPGA's can be used to speed up the communications processing on board the vehicle and can be directly interfaced with the vehicle's buses, for instance the control bus or CANBUS. 
\item \textbf{Speeding up and enhancing the efficiency of the side-channels}. This can be achieved by adding extra transmitters/receivers that modulate separate information in parallel; for example, by using multiple brake lights, segmenting brake lights, using scrolling light patterns, using non human-visible light sources; and through using multiple ultrasonic transmitters, MIMO techniques and variable frequency transducers. Detector advances will also contribute significantly e.g. the fast CMOS sensor technology \cite{goodcmos} mentioned previously with a 15-Mbps/pixel data rate. 
\item \textbf{Validating and enhancing the security of both blockchains and blockchain public key infrastructures}. The cryptographic algorithms behind blockchains are robust and proven to be sufficiently secure. However, the security of the implementation of the blockchain network, and impact on the  original security properties of the underlying blockchain system when non-blockchain related information is introduced, remains unclear (in our example cryptographic information for securing vehicular communications is inserted). 
\item \textbf{Practical characterisation of visual and Auditory Sensing Ranges}.  Drive testing is required to provide input for establishing the real-world effective ranges of each sidechannel, and for determining the upper and lower bounds on detection and communication ranges. 
\end{itemize}

\section{Conclusion}

Vehicular networking and communication is becoming more common, in part driven by autonomous vehicles and the need to increase occupant safety e.g. by reducing poor human-in-the-loop reaction times in critical control situations. Of particular concern to manufacturers, regulators and owners are the risks of electronic attack and compromise of operational vehicles via the vehicular communication infrastructures, e.g. a large scale, localised interference attack on radio frequency communication channels. This paper proposes a novel secure inter-vehicle communication system using side-channels (visual light and ultrasonic audio) that is highly robust to interference and attack.  The scheme verifies the location and identity of the vehicle being communicated with, and incorporates that vehicle identifier in the cryptographic setup exchanges. An original key establishment handshake protocol is also presented. This is primarily based on TLS 1.2 and it limits the throughput requirement to 176 bits on the side-channels when establishing symmetric encryption and message authentication keys and when verifying a vehicle's certificate with accepted certification authorities. The system exploits both physical side-channels, and employs a blockchain public key infrastructure for interoperability between untrusted vehicles and manufacturers. The side-channels provide enhanced physical (directional) security to the transmissions, and are directly useful in passing messages between vehicles and, for instance, in maintaining the inter-vehicle distance in a vehicular platoon.






%

\end{document}